\documentclass[twocolumn,amsmath,amssymb,floatfix,prl]{revtex4}

\usepackage{graphicx}
\usepackage{amsmath}
\usepackage{bm}

\begin{document}

\title{Direct mapping of the formation of a persistent spin helix}

\author{M. P. Walser$^1$, C. Reichl$^2$, W. Wegscheider$^2$, G. Salis$^{1,}$\footnote[1]{Electronic address: gsa@zurich.ibm.com}}
\affiliation{$^1$IBM Research--Zurich, S\"aumerstrasse 4, 8803
R\"uschlikon, Switzerland}
\affiliation{$^2$Solid State Physics
Laboratory, ETH Zurich, 8093 Zurich, Switzerland}

\date{25. June 2012}

\begin{abstract}
The spin-orbit interaction (SOI) in zincblende semiconductor quantum wells can be set to a symmetry point, in which spin decay is strongly suppressed for a helical spin mode. Signatures of such a persistent spin helix (PSH) have been probed using the transient spin grating technique, but it has not yet been possible to observe the formation and the helical nature of a PSH. Here we directly map the diffusive evolution of a local spin excitation into a helical spin mode by a time- and spatially resolved magneto-optical Kerr rotation technique.  Depending on its in-plane direction, an external magnetic field interacts differently with the spin mode and either highlights its helical nature or destroys the SU(2) symmetry of the SOI and thus decreases the spin lifetime. All relevant SOI parameters are experimentally determined and confirmed with a numerical simulation of spin diffusion in the presence of SOI.
\end{abstract}

\maketitle

Conduction-band electrons in semiconductors experience SOI from intrinsic~\cite{Winkler2003} and extrinsic sources, leading to spin dephasing, current-induced spin polarization and spin Hall effects~\cite{Dyakonov2008}. These physical mechanisms are of great fundamental and technological interest, recently also in the context of topolocial insulators~\cite{Koenig2007} and Majorana fermions~\cite{Sau2010, Mourik2012}. Intrinsic SOI arises from an inversion asymmetry of the bulk crystal (Dresselhaus term) and of the grown layer structure (Rashba term). In a quantum well (QW), these two components can be tailored by means of the confinement potential~\cite{Koralek2009}, and the Rashba SOI can be externally tuned by using gate electrodes~\cite{Nitta1997,Studer2009b}. In general, SOI leads to precession of electron spins. In the diffusive limit, in which the scattering length is much smaller than the spin-orbit (SO) length $\lambda_\textrm{SO}$, a random walk of the spins on the Bloch sphere will dephase a non-equilibrium spin polarization~\cite{Dyakonov1972}.

\begin{figure*}[ht] 
\includegraphics{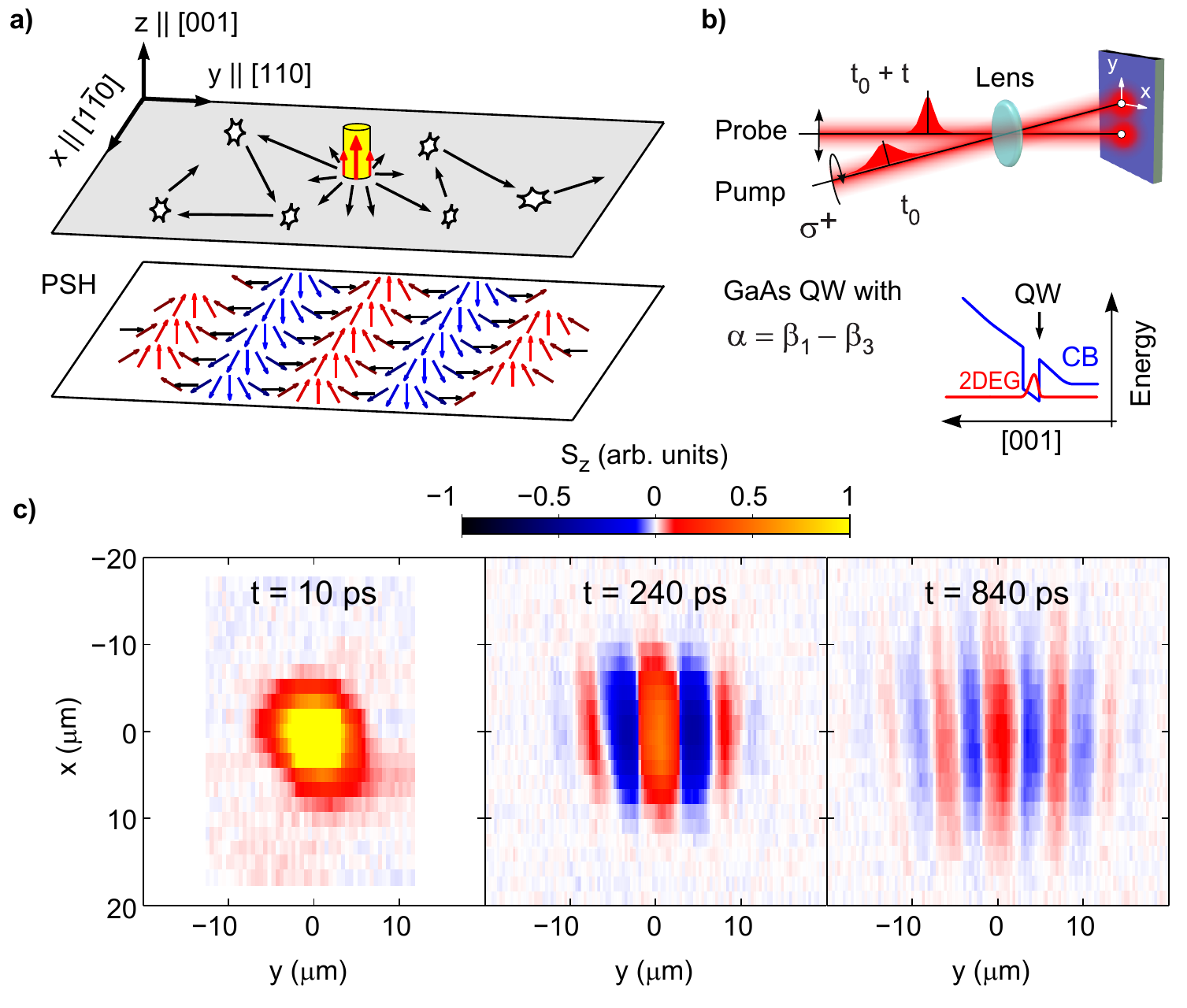}
\caption{\label{fig:fig1} \textbf{Direct mapping of the persistent spin helix formation.}
\textbf{(a)}, Diffusive expansion of a local spin excitation (top), where the spin polarization evolves into a PSH mode (bottom). The arrows and the colors indicate the direction of $\mathbf{S}$ and the magnitude of $S_z$ respectively.
\textbf{(b)}, Schematic of time-resolved Kerr rotation microscopy (see methods) and profile of conduction-band (CB) energy of the 12-nm-wide GaAs/AlGaAs QW sample investigated.
\textbf{(c)}, Experimental observation of the PSH. Spatial maps of $S_z(x,y)$ are shown for three different times $t$.}
\end{figure*}

Of special interest is the situation in a two-dimensional electron gas (2DEG) with balanced Rashba and Dresselhaus contributions~\cite{Schliemann2003,Bernevig2006,Koralek2009,Wunderlich2010,Duckheim2007}. There, the SOI attains SU(2) symmetry and the spin polarization of a helical mode is preserved. The reason for this conservation of the spin polarization is a unidirectional effective SO magnetic field $\mathbf{B}_\textrm{SO}$, which depends linearly on the component of the electron momentum along a specific in-plane direction. This causes the precession angle of a moving electron to vary linearly with the distance traveled along that direction, irrespective of whether the electron path is ballistic or diffusive~\cite{Schliemann2003,Bernevig2006}. In such a situation, a local spin excitation is predicted to evolve into a helical spin mode termed PSH [Fig.~\ref{fig:fig1}(a)]. Transient spin grating measurements~\cite{Koralek2009} showed that a spin excitation with a spatially modulated out-of-plane spin component decays with two characteristic lifetimes that correspond to two superposed spin modes of opposite helicity.

Here we directly measure the diffusive evolution of a local spin excitation into a PSH by time-resolved Kerr rotation microscopy [Fig.~\ref{fig:fig1}(b)]. We employ a pump-probe approach, in which a circularly polarized pump pulse excites electrons into the conduction band of a (001)-grown GaAs/AlGaAs QW with their spins polarized along $z||[001]$. The out-of-plane spin polarization $S_z$ is then measured by a probe pulse delayed by a time $t$, using the polar magneto-optical Kerr effect. The position of the incident pump beam is scanned to record the spatial spin distribution $S_z(x,y)$ at time $t$. We define $x$ along the [1$\overline1$0] and $y$ along the $[110]$ direction.
The SOI of the 2DEG is tuned close to the SU(2) symmetry point, $|\alpha|\approx|\beta_1-\beta_3|$, by controlling the Rashba ($\alpha$), the linear Dresselhaus ($\beta_1$) and the cubic Dresselhaus ($\beta_3$) SO coupling coefficient via asymmetric modulation doping on the two sides of the QW.

The experimental observation of the PSH is exemplified by three maps of $S_z(x,y)$ recorded at different $t$ [Fig.~\ref{fig:fig1}(c)]. The first map at $t=10$\,ps still shows the local excitation of $S_z>0$ centered at $x=y=0$. Because of the initially rapid spin diffusion, the Gaussian shape of $S_z(x,y)$ is already broader than the size of the focused pump-laser spot. Spins further diffuse in the $(x,y$) plane, but the second and the third map (recorded at $t=240$ and 840\,ps) in addition feature alternating stripes of $S_z(x,y)>0$ and $S_z(x,y)<0$  caused by spin precession about $\mathbf{B}_\textrm{SO}$. To explain this unidirectional oscillation along the $y$-direction, $\mathbf B_\textrm{SO}$ must be more strongly correlated with $k_y$ than with $k_x$ ($k_x$ and $k_y$ are the components of the electron wave vector $\mathbf k$). With our definition of $\alpha$ and $\beta$ and from the symmetry of $\mathbf B_\textrm{SO}$ (see supplementary information), it follows that $\alpha$ and $(\beta_1-\beta_3)$ must have the same signs and that therefore the $x$-component of $\mathbf B_\textrm{SO}$ is much larger than the $y$-component. For opposite signs of $\alpha$ and $(\beta_1-\beta_3)$, the PSH would oscillate along the $x$-direction and the $y$-component of $\mathbf B_\textrm{SO}$ would be larger.

\begin{figure*}[ht]
\includegraphics{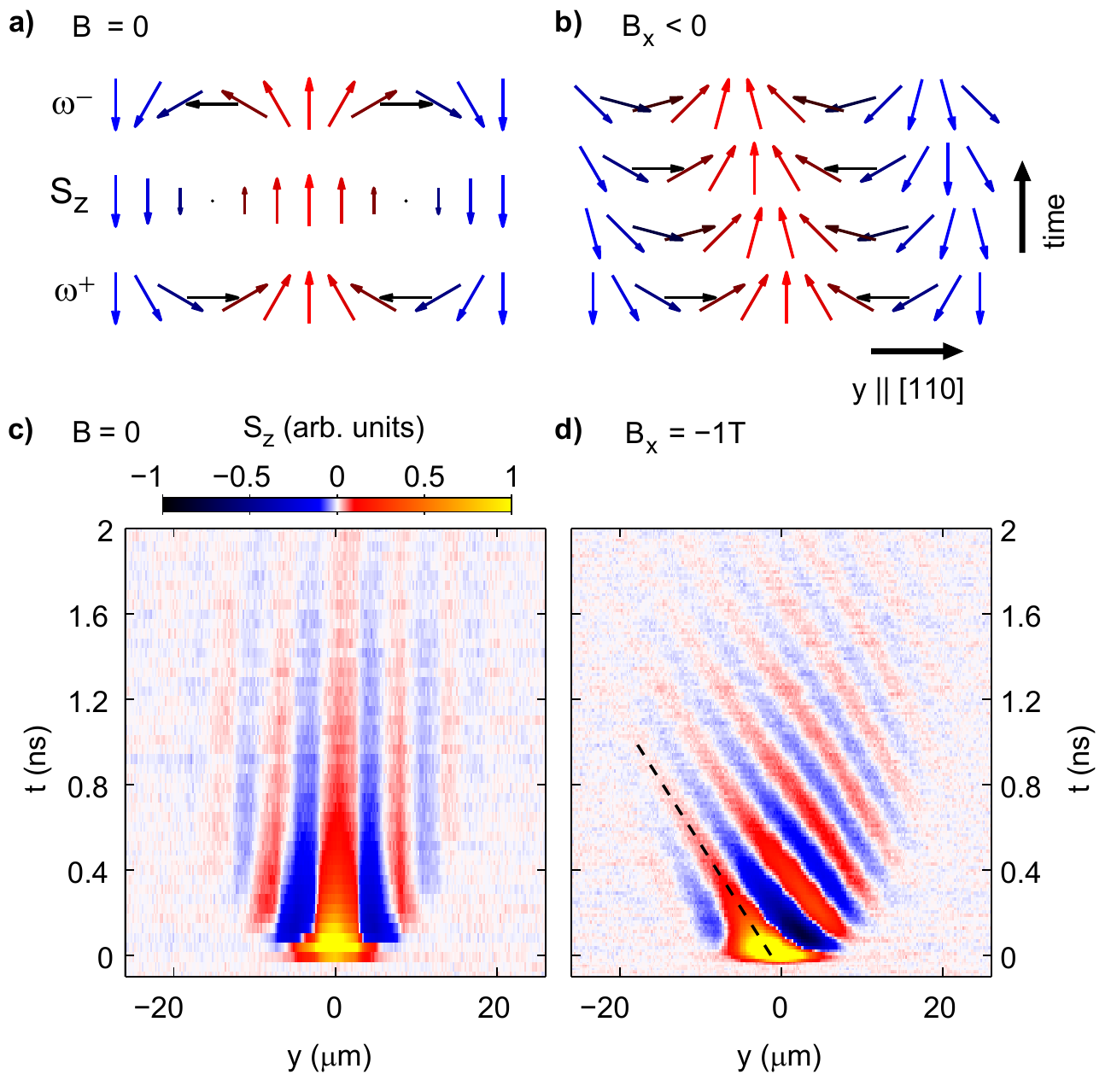}
\caption{\label{fig:fig2} \textbf{Helical spin modes and the PSH.}
\textbf{(a)}, The spins of the two helical spin modes $\omega^+$ and $\omega^-$ rotate with opposite helicity as the electron moves along $y$. The $S_z(y)$ of the two modes is the same.
\textbf{(b)}, Spins precess about an external magnetic field $B_x$ such that the $\omega^{+}$ mode is shifted towards $-y$ if $B_x<0$ is applied.
\textbf{(c), (d)}, Maps of $S_z(y,t)$ show the time evolution of the PSH for $B_x=0$ (c) and $B_x=-1$\,T (d). $S_z(y)$ oscillates in both cases, but for $B_x=-1\,$T, the position of constant phase, $y_0$, is shifted with $t$ (marked by the dashed line).}
\end{figure*}

The formation of the PSH is best illustrated if spin dynamics is tracked in space and time.
For that purpose, we position the pump pulse at $x=0$ and scan it along the $y$-direction. A collection of such line scans $S_z(y,t)$ recorded at various $t$ is shown as color-scale plot in Fig.~\ref{fig:fig2}(c). Starting with $\mathbf{S}\parallel z$, the excited spin distribution expands along $\pm y$, and thereby $S_z$ starts to oscillate with $y$. As we will discuss in the following, this oscillation is indeed the footprint of a helical spin mode.

A cartoon of helical spin modes is shown in Fig.~\ref{fig:fig2}(a). The spin polarization rotates by an angle that depends linearly on the position $y$, with the direction of rotation defining a helicity ($\omega^{+}$ or $\omega^{-}$). Therefore, not only $S_z$ but also $S_y$ oscillates with $y$ [Fig.~\ref{fig:fig2}(a)]. The helicity of the emerging spin mode depends on the sign of the cross-correlation $\langle B_{\textrm{SO},x} \: k_y\rangle$, i.e., on whether $B_{\textrm{SO},x}$ is positive or negative for $k_y>0$, and is given by the absolute sign of $\alpha+\beta_1-\beta_3$ (see supplementary information). The helicity cannot be directly determined from $S_z(y)$, as for both modes, $S_z$ oscillates in the same way [Fig.~\ref{fig:fig2}(a)].  This is especially true for experiments employing the transient spin-grating method, in which as an additional complication, spin waves with both helicities are excited simultaneously~\cite{Koralek2009}.

We uncover the helical nature of the measured spin mode by rotating in-plane spin components $S_y$ out of plane with the help of an external magnetic field $\mathbf{B}\parallel x$ [Fig.~\ref{fig:fig2}(b)]. The effect of $B_x$ on $S_z(y,t)$ is shown in Fig.~\ref{fig:fig2}(d). $S_z(y,t)$ still oscillates with $y$ -- indicating that the PSH is preserved -- but the phase of the oscillation now shifts with $t$  (dashed line). This shift can be understood from the spin precession about $B_x$, as shown in Fig.~\ref{fig:fig2}(b): For an $\omega^{+}$ mode and $B_x<0$, the position $y_0$ of equal phase shifts toward $-y$ with increasing $t$. As we observe the same direction in the measurement [Fig.~\ref{fig:fig2}(d)], we conclude that our sample supports an $\omega^{+}$ mode, which means that $\alpha+\beta_1-\beta_3$ must be positive. With this, we have directly determined the sign of the cross-correlation $\langle B_{\textrm{SO},x} \: k_y\rangle$.

Figure~\ref{fig:fig2}(d) also shows that for all positions $y$, $S_z(y,t)$ oscillates in $t$ with the same frequency $\nu$. Therefore, $S_z(y,t)$ represents a collective precession of the $\omega^{+}$ mode about $B_x$. According to $\nu=|g\mu_B B_x/h|$, we determine the electron g-factor $g$ to equal $-0.17$ (based on the QW thickness, we assume $g<0$; $\mu_B$ is the Bohr magneton, and $h=2\pi\hbar$ Planck's constant).

\begin{figure*}[ht]
\includegraphics[scale=0.8]{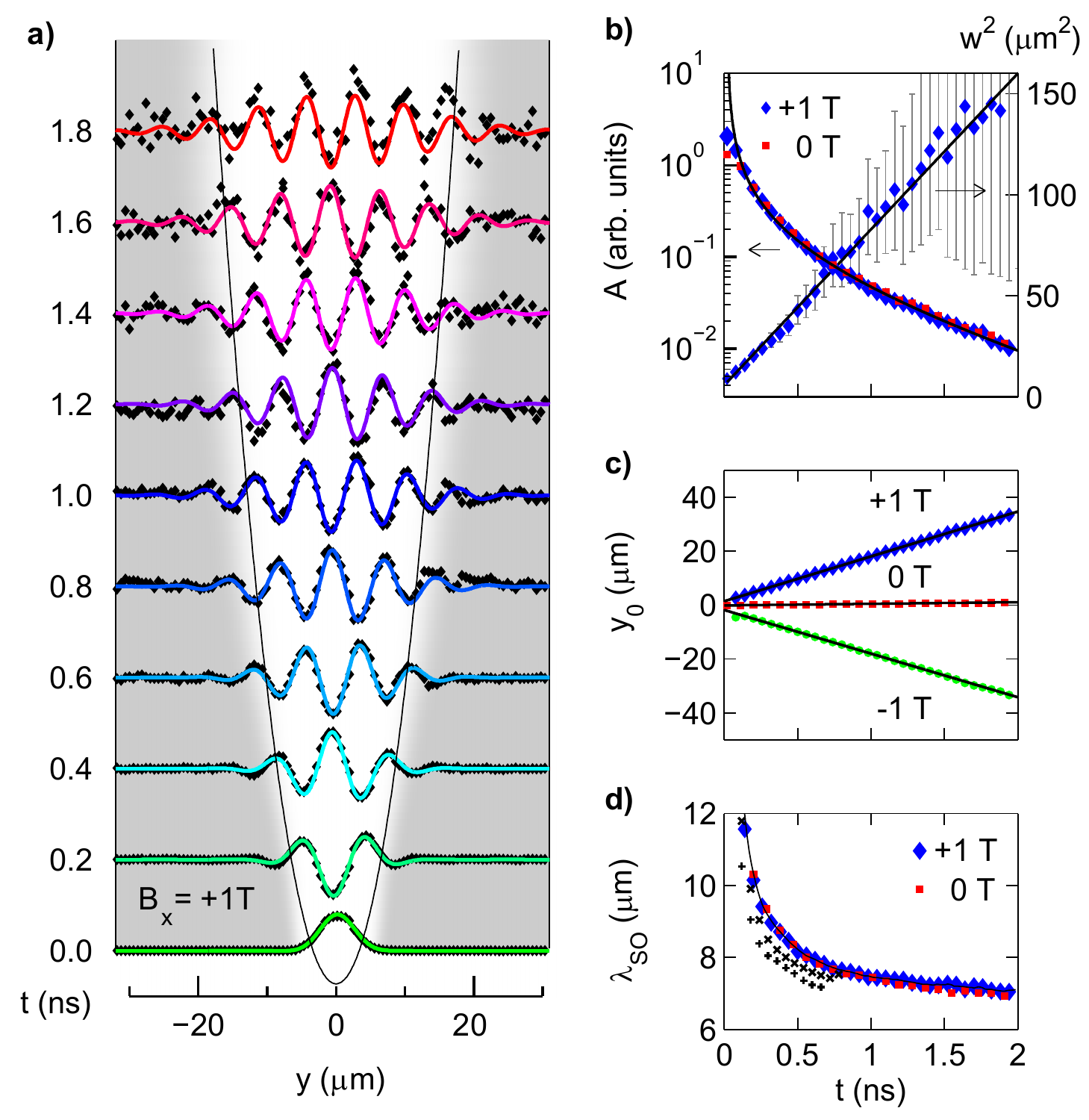}
\caption{\label{fig:fig3} \textbf{Spin diffusion and SOI characterization.} \textbf{(a)}, Normalized line scans $S_z(y)$ for various $t$. The experimental data (symbols) are fit to $A\exp(-y^2/4w^2)\times\cos\big(2\pi(y-y_0)/\lambda_{\textrm{SO}}\big)$ (solid lines). The parabola follows $\pm2\sqrt{D_s t}$. \textbf{(b)}, Symbols: Amplitude $A$ and squared width $w^2$ of the Gaussian envelope. Solid lines: fits to $(t-t_0)^{-1}\exp(-t/\tau_s)$ and $D_s t$, respectively. Error bars: Uncertainty of $w^2$ in a single fit. \textbf{(c)}, Shift $y_0$ of the spatial oscillation for $-1$, 0 and $1\,$T.  \textbf{(d)}, Symbols: PSH period $\lambda_{\textrm{SO}}$ for 0 and 1~T. Solid line: smoothed data at 1~T. ($\scriptstyle \times$) and ($\scriptstyle +$): fits to data at three- and nine-fold lower pump intensities.}
\end{figure*}

We now quantitatively analyze $S_z(y,t)$ to extract information on the SOI and on spin diffusion.
Figure~\ref{fig:fig3}(a) shows $S_z(y)$ for different $t$ with $B_x=+1$\,T. The experimental data (symbols) are fit to the product of a Gaussian, $A\exp(-y^2/4w^2)$, and a cosine function, $\cos\big(2\pi(y-y_0)/\lambda_{\textrm{SO}}\big)$ (solid lines). The Gaussian envelope describes the solution of the spin diffusion equation, for which $w^2=D_s t$ ($D_s$ is the spin diffusion constant). Spin dephasing is included in the time dependence of the amplitude $A(t)$.

A linear fit of $w^2$ to $D_s t$ provides a direct measure of $D_s$ [Fig.~\ref{fig:fig3}(b)], and we obtain $D_s=(385\pm15)$\,cm$^2$s$^{-1}$. Note that our experiment uses the spin as a label of the electrons and therefore tracks the spin and not the charge diffusion. The sensitivity of spin diffusion to electron-electron interactions~\cite{Amico2000,Yang2011} explains the ten-fold smaller value of $D_s$ as compared to the charge diffusion constant $D\approx 4000\,$cm$^2$s$^{-1}$ (as calculated for the measured electron mobility $\mu\approx22\,$m$^2$/Vs and a sheet carrier density $n_s=5\times 10^{15}\,$m$^{-2}$).

Figure~\ref{fig:fig3}(b) shows the decay of $A(t)$ with $t$.
The diffusive expansion of the excited spins into an area proportional to $D_s t$ decreases $S_z$ proportional to $(D_s t)^{-1}$. Additional spin decay is mainly induced by deviations from perfect SU(2) symmetry~\cite{Yang2010} and can be described by an exponential decay proportional to $\exp{(-t /\tau_s)}$, where $\tau_s$ is the spin lifetime. $A(t)$ is therefore fit to $(t-t_0)^{-1}\exp(-t/\tau_s)$, yielding $\tau_s=(1.1\pm0.1)\,$ns. This is about 30 times longer than the Dyakonov--Perel spin dephasing time calculated from the measured $D_s$ and SOI strength, in agreement with a spin decay time $\tau_z\approx35$\,ps obtained for 20-$\mu$m-wide laser spots, where the spatial correlations of the PSH are averaged out (see supplementary information).

The measured $\tau_s$ is limited by the two SU(2)-breaking contributions, namely, the cubic Dresselhaus SOI $\beta_3$ and the imbalanced SOI $|\alpha| \neq |\beta_1-\beta_3|$. It can be shown that  (supplementary information)
\begin{equation}\label{eq1}
\tau_s^{-1} \approx 2D_s {\frac{m^2}{\hbar^4}} \big(3 \beta_3^2 + (\alpha-\beta_1+\beta_3)^2\big),
\end{equation}
where $m=6.1\times10^{-32}$\,kg is the effective electron mass. From the measured $\tau_s$ and $D_s$, we find $3\beta_3^2+(\alpha-\beta_1+\beta_3)^2=1.7\times10^{-26}\,$eV$^2$m$^2$. This relation restricts $\beta_3$ to an upper limit of $0.77\times 10^{-13}$\,eVm, which would be reached for $\alpha=\beta_1-\beta_3$. From the measured $\tau_s$ alone, it is not possible to differentiate cubic Dresselhaus contributions from imbalanced SOI. However, as we will discuss later, the latter can be separately determined from a small asymmetry in $S_z(x,t)$ maps that appears if $\mathbf{B}$ is applied along $y$. This will allow us to quantitatively describe all SO coefficients.

The fitted PSH period $\lambda_{\textrm{SO}}$ is shown in Fig.~\ref{fig:fig3}(d). A decrease from $\approx10\,\mu$m at $t=200\,$ps to $\approx7.3\,\mu$m at 1.5\,ns is well represented in several periods of $S_z(y)$ and must be related to a continuous change of $\lambda_\textrm{SO}$ with $t$. From the relation $\lambda_\textrm{SO}=\pi \hbar^2 m^{-1} (\alpha + \beta_1 - \beta_3)^{-1}$, we determine that $|\alpha + \beta_1 - \beta_3|$ increases from 3.5 to 4.9$\times10^{-13}\,$eVm during this time. Starting with a broad positive peak in $S_z(y)$ at $t=10$\,ps, the spin helix continuously adapts to the decreasing $\lambda_\textrm{SO}$. The initially weaker SOI is most likely induced by the photo-excited charge carriers that recombine with time, and is possibly also affected by cooling of hot electrons. The SO coefficients are sensitive to the screening of the confinement potential and modifications of the Fermi energy: Both an increase of $\beta_3$ with charge density and a reduction of $\alpha$ and $\beta_1$ with screening could explain the observed decrease of $\lambda_\textrm{SO}$ with $t$. Supporting this interpretation, measurements at lower pump intensities yield an initially smaller  $\lambda_\textrm{SO}$ [Fig.~\ref{fig:fig3}(d)]).

$\lambda_{\textrm{SO}}$ is found to be independent of $B_x$ [Fig.~\ref{fig:fig3}(d)]. Together with the insensitivity of $A(t)$ on $B_x$, this demonstrates the decoupled influence of the Zeeman and the SO energy on the electron spins in the case where a magnetic field is applied along the unidirectional $\mathbf B_\textrm{SO}$.


\begin{figure}[ht]
\includegraphics[width=80mm]{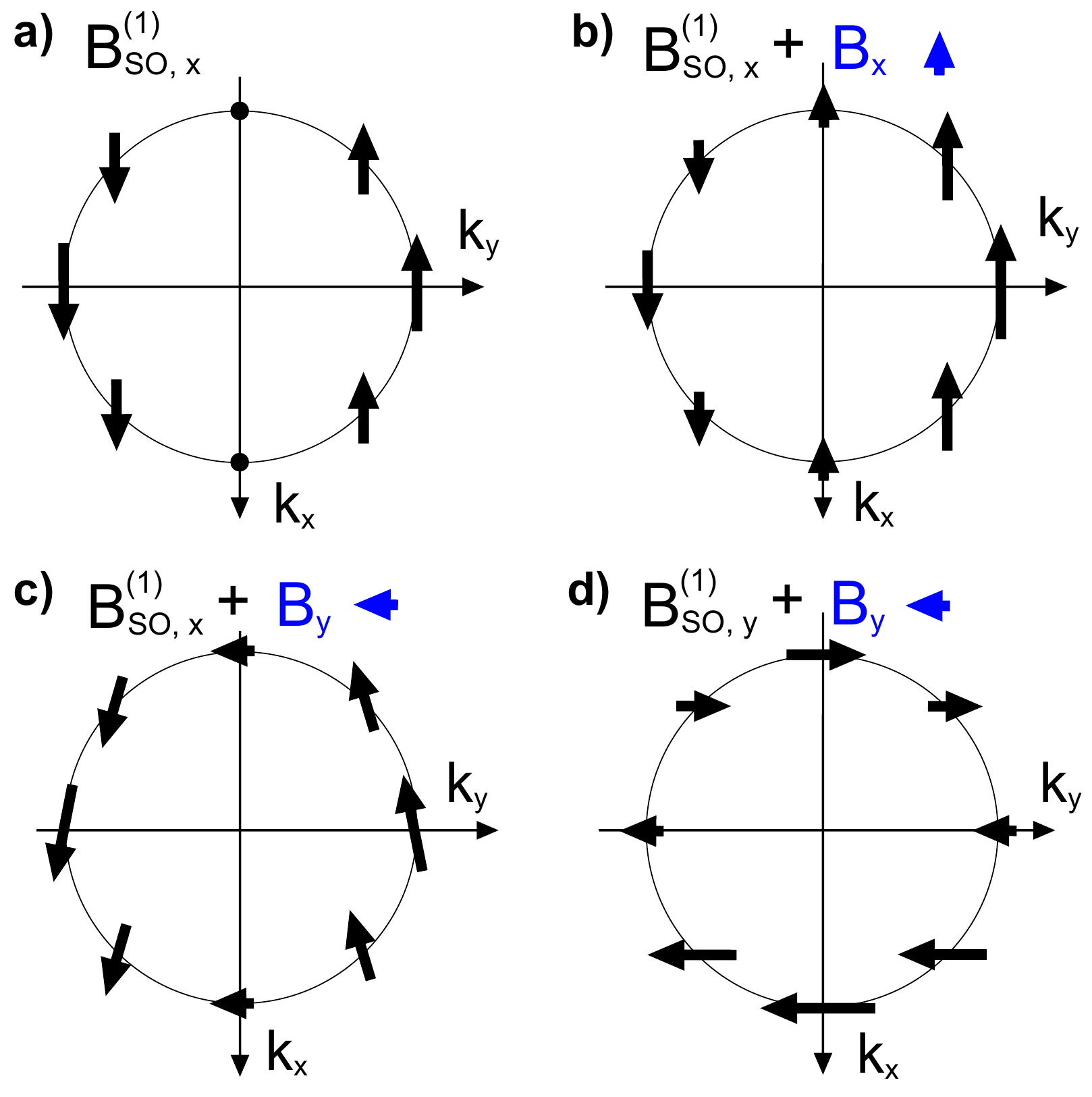}
\caption{\label{fig:fig4} \textbf{Dependence of the total magnetic field $B_\textrm{tot}$ on k.} \textbf{(a)}, The $x$-component of $\mathbf{B}_{\textrm{SO}}^{(1)}$ is shown (arrows) as a function of $\mathbf k$ for $\alpha>0$, $\beta_1-\beta_3>0$ and $g<0$. \textbf{(b)}, If an external magnetic field $\mathbf{B}$ is applied along $x$, $|B_{\textrm{tot}}|$ is different for $k_y>0$ and $k_y<0$. The arrows represent the size and direction of $B_{\textrm{SO},x}^{(1)}+B_x$. \textbf{(c)}, For $\mathbf{B}\parallel y$, $|B_{\textrm{tot}}|$ is the same for $\pm k_y$. \textbf{(d)}, Away from the SU(2) symmetry point ($\alpha\ne\beta_1-\beta_3$), the $y$-component of $\mathbf{B}_{\textrm{SO}}^{(1)}$ is non-zero, and the superposition with $B_y$ leads to a different $B_{\textrm{tot}}$ for $k_x>0$ than for $k_x<0$. This allows us to determine $\alpha-\beta_1+\beta_3$.}
\end{figure}

To understand this decoupling better, it is instructive to plot the directional dependence of $\mathbf{B}_{\textrm{SO}}$ on $\mathbf{k}=k(\cos\theta,\sin\theta)$.
For (001)-grown QWs, $\mathbf{B}_{\textrm{SO}}$ is in the $(x,y)$-plane for all $\mathbf{k}$.
It can be written as the sum of two terms, $\mathbf{B}_{\textrm{SO}}^{(1)}$ and $\mathbf{B}_{\textrm {SO}}^{(3)}$. The former is responsible for the PSH formation, whereas the latter leads to spin dephasing (see supplementary information). The $x$- and $y$-components of $\mathbf{B}_{\textrm{SO}}^{(1)}$ are proportional to $k_y(\alpha+\beta_1-\beta_3)$ and $-k_x(\alpha-\beta_1+\beta_3)$, respectively. In our case, $|\alpha+\beta_1-\beta_3| \gg |\alpha-\beta_1+\beta_3|$, which means that $B^{(1)}_{\textrm{SO},x}$ drives the PSH [Fig.~\ref{fig:fig4}(a)], whereas a remaining $B^{(1)}_{\textrm{SO},y}$ breaks the SU(2) symmetry and leads to spin dephasing. In the following discussion, we omit the superscript and mean $\mathbf B^{(1)}_\textrm{SO}$ when we write $\mathbf B_\textrm{SO}$.

As shown in Fig.~\ref{fig:fig4}(b), $B_x<0$ superposes with $B_{\textrm{SO},x}$ such that the total field, $B_\textrm {tot}=|\mathbf B + \mathbf B_\textrm{SO}|$, is larger for $k_y>0$ than for $k_y<0$. Translating the momentum $\hbar k_y$ into a position $y$ using $y=\hbar k_y t/m$, this is exactly what is seen in the measurement at $B_x=-1$\,T in Fig.~\ref{fig:fig2}(d):
The total precession frequency $\nu$ of an electron with momentum $\hbar k_y$ is determined by the sum of the Zeeman splitting, $g \mu_\textrm{B} B_x$, and the SO splitting, $2k_y(\alpha + \beta_1 - \beta_3)$. Electrons with $k_y=0$ remain at $y=0$ and precess with $\nu=|g\mu_B B_x/h|$. There is a path $y_0(t)$ [dashed line in Fig.~\ref{fig:fig2}(d)] on which Zeeman and SO energies cancel each other and consequently the spins do not precess. This path is characterized by
\begin{equation}
\label{eq2}
\partial y_0/\partial t = -\hbar g \mu_{\textrm B} B_x / [2m(\alpha+\beta_1-\beta_3)].
\end{equation}
In agreement with this equation, the fitted $y_0$ increases linearly with $t$ and $\partial y_0/\partial t$ changes sign with $B_x$ [Fig.~\ref{fig:fig3}(c)]. Inserting the measured $\partial y_0/\partial t$ and $g$ into Eq.~(\ref{eq2}), we obtain $\alpha+\beta_1-\beta_3=4.8\times10^{-13}$\,eVm, consistent with the value obtained from $\lambda_\textrm{SO}$, but here the sign is directly determined by the sign of $\partial y_0/\partial t$.


\begin{figure*}[ht]
\includegraphics{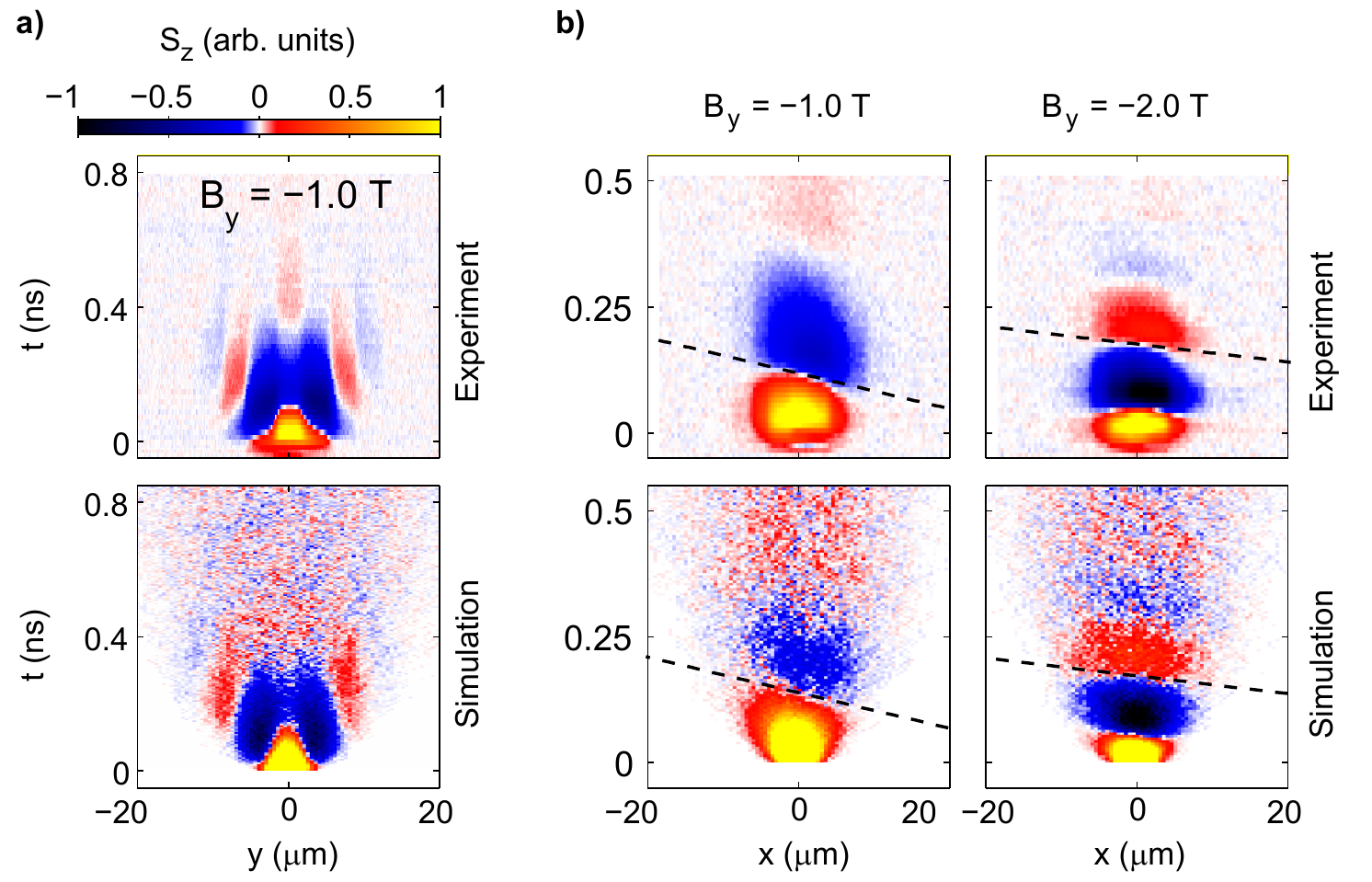}
\caption{\label{fig:fig5} \textbf{Interplay of the PSH with an external magnetic field.} \textbf{(a)}, Experimentally measured and numerically simulated maps of $S_z(y,t)$ for $B_y=-1$\,T. The formation of a helical spin mode is challenged by simultaneous spin precession about $B_y$. \textbf{(b)}, Maps of $S_z(x,t)$ recorded at $B_y=-1$ and $-2$\,T show the spin precession about $B_y$ as well as an asymmetry of the precession phase with $\pm x$ (dashed lines), which is attributed to a slight detuning from the SU(2) symmetry point.}
\end{figure*}

We now investigate how close the SOI in our QW is tuned to the SU(2) symmetry point. For that purpose we apply $\mathbf B$ along $y$. Figure~\ref{fig:fig4}(c) shows that then, $B_\textrm{tot}$ is not unidirectional anymore. Even though a map of $S_z(y,t)$ at $B_y=-1\,$T [Fig.~\ref{fig:fig5}(a)] still indicates that a helical mode evolves for $t<400\,$ps, $S_z$ quickly decays at longer $t$. In addition, an oscillation with $t$ is seen at $(x,y)=(0,0)$. This is related to the precession of $\mathbf{S}$ about $B_y$. On the other hand, the superposition of $B_y$ and $B_{\textrm{SO},y}$ [Fig.~\ref{fig:fig4}(d)] leads to an asymmetry for opposite signs of $k_x$, which can be observed in maps of $S_z(x,t)$ [Fig.~\ref{fig:fig5}(b)]. We find a small tilt of the positions $x_0(t)$ of constant spin precession phase. In analogy to Eq.~(\ref{eq2}), the tilt is given by
\begin{equation}
\label{eq3}
\partial x_0/\partial t = \hbar g \mu_{\textrm B} B_y / [2m(\alpha-\beta_1+\beta_3)].
\end{equation}
$\partial x_0/\partial t$ is a measure of the detuning from balanced SOI. We obtain $\partial x_0/\partial t\approx\,-280\,\mu$m/ps for $B_y=-1\,$T, and about twice that value for $B_y=-2$\,T [dashed lines in Fig.~\ref{fig:fig5}(b)]. From this, it follows that $\alpha-\beta_1+\beta_3\approx-0.3\times10^{-13}$\,eVm, where the sign is directly determined by the sign of $\partial x_0/\partial t$.

We can now derive the size of all SO coefficients in our sample. From the known sum and difference of $\alpha$ and $\beta_1-\beta_3$, respectively, we find $\alpha=( 1.6-2.3)\times10^{-13}\,$eVm and $\beta_1-\beta_3=(1.9-2.6)\times10^{-13}\,$eVm. The cubic Dresselhaus SOI $\beta_3$ is estimated from Eq.~(\ref{eq1}) based on the observed $\tau_s$, yielding $\beta_3\approx0.7\times10^{-13}\,$eVm. We then obtain $\beta_1\approx (2.6-3.3)\times10^{-13}$\,eVm, which is in excellent agreement with results from a similar QW structure~\cite{Koralek2009}. Using the theoretical expression $\beta_1=-\gamma \langle k_z^2 \rangle$~\cite{Winkler2003}, with $\langle k_z^2 \rangle=3.7\!\times\!10^{16}$~m$^{-2}$ as obtained by solving the one-dimensional Poisson and Schr\"odinger equations, we determine the Dresselhaus coupling parameter $\gamma\approx -9\,$eV\AA$^3$. This is in agreement with previous work where $\gamma$ was found to be in the range of $-4$ to $-8\,$eV\AA$^3$~\cite{Studer2009b}. Inserting $\gamma=-9\,$eV\AA$^3$ into $\beta_3=-\tfrac{1}{2}\gamma \pi n_s$, with $n_s=5\times 10^{15}\,$m$^{-2}$, we find $\beta_3=0.7\times10^{-13}$\,eVm, exactly the same value as obtained from $\tau_s$. This indicates that the PSH decay is well described by Eq.~(\ref{eq1}).
The Rashba coefficient $\alpha$ can be related to the electric field $E_{\textrm {QW}}$ in the QW by $\alpha=r_\textrm{QW} E_\textrm{QW}$, defining a proportionality constant $\alpha=r_\textrm{QW}$. From the calculated conduction-band profile, we estimate $E_{\textrm {QW}}= 4-6\times 10^6$V/m. Using $\alpha\approx 2.3\times10^{-13}$\,eVm, we obtain $r_{\textrm {QW}}= 4-6$\,e\AA$^2$, in good agreement with theoretical prediction~\cite{Winkler2003} and experiment~\cite{Koralek2009}.

To crosscheck the consistency of our experimentally determined SO coefficients, we simulate spin diffusion in the presence of SOI using a Monte Carlo approach that combines semiclassical spin dynamics and diffusion (see numerical methods). The simulation uses the experimentally determined values for $\alpha_1$, $\beta_1$, $\beta_3$ and $D_s$, and reproduces the spin dynamics in the combined field of $\mathbf{B}$ and $\mathbf{B_{\textrm{SO}}}$ remarkably well [see Figs.~\ref{fig:fig5}(a) and (b)]. It also confirms the analytical relation of Eq.~(\ref{eq2}) [dashed lines in Fig.~\ref{fig:fig5}(b)].

\begin{figure}[ht]
\includegraphics[width=85mm]{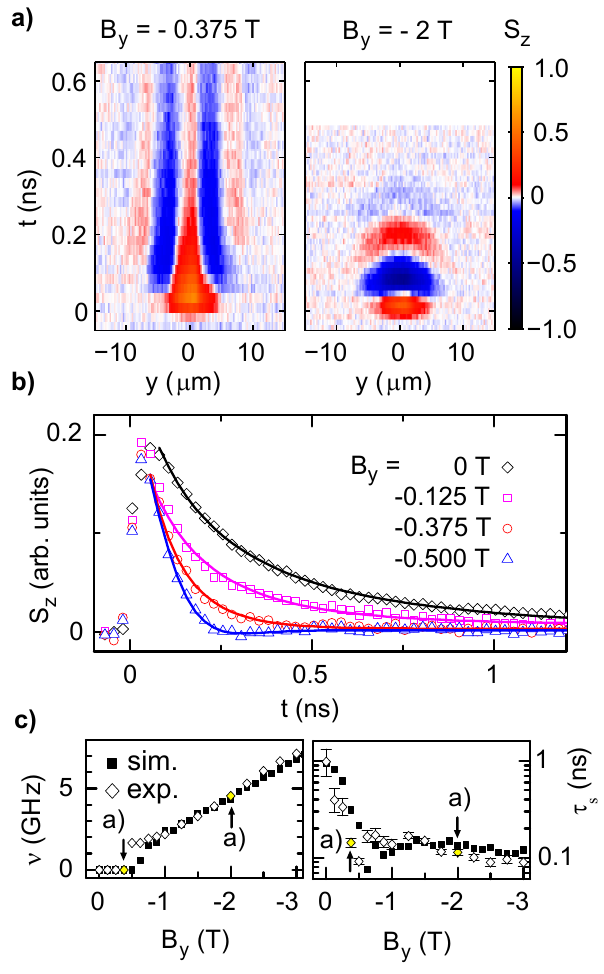}
\caption{\label{fig:fig6} \textbf{Detuning from the PSH regime.} \textbf{(a)}, Left: Maps of $S_z(y,t)$ for $B_y=-0.375$\,T. The PSH is still visible, but the spin lifetime has decreased to a few 100$\,$ps. Right: Maps of $S_z(y,t)$ for $B_y=-2\,$T. Instead of the formation of a PSH, a spin precession about $B_y$ is seen. \textbf{(b)}, Time traces $S_z(t)$ exhibit no spin precession for $|B_y|\leq 0.375$\,mT, and $\tau_s$ decreases rapidly with $|B_y|$. \textbf{(c)}, Precession frequency $\nu$ and spin lifetime $\tau_s$ versus $B_y$. Open symbols: Experimental data. Filled symbols: Numerical simulation.}
\end{figure}

Finally we want to discuss the transition from a PSH-dominated regime to a regime in which the external magnetic field dominates. Figure~\ref{fig:fig6}(a) shows two maps of $S_z(y,t)$ recorded at $B_y=-0.375$ and $-2$\,T.
At $B_y=-0.375$\,T, the formation of a helical mode is readily identifiable, but $\tau_s$ is significantly lower than for 0\,T [see Fig.~\ref{fig:fig2}(c)]. At $-2$\,T, no signature of a helical mode is observed and spins precess about $B_y$ with $\nu=|g\mu_B B_y/h|$.
To determine $\nu$ and $\tau_s$ versus $B_y$, the time traces $S_z(t)$ at $(x,y)=(0,0)$ [Fig.~\ref{fig:fig6}(b)] are fit to $(t-t_0)^{-1}\exp(-t/\tau_s)\cos(2\pi\nu t)$. Starting from the PSH at 0$\,$T, $\tau_s$ rapidly decreases with  increasing $|B_y|$ [Fig.~\ref{fig:fig6}(c)]. Interestingly, spin precession about $\mathbf{B}$ is suppressed in the regime of $|B_y|\leq0.5\,$T. Even though at $|B_y|=0.5\,$T the Zeeman spin splitting (5\,$\mu$eV) is more than one order of magnitude smaller than the SO spin splitting (up to 170$\,\mu$eV at the Fermi energy), the symmetry-breaking induced by $B_y$ dramatically decreases $\tau_s$ from 1$\,$ns to below 100$\,$ps [Fig.~\ref{fig:fig6}(c)]. In an intermediate regime of $0.5$\,T$<|B_y|<1.5$\,T, the helical spin mode coexists with spin precession about $\mathbf{B}$, and $\tau_s$ recovers from its minimum value. For $|B_y|>1.5\,$T, $\tau_s$ monotonically decreases. In the regime where the Zeeman energy approaches the SOI energy, we expect that $\tau_s$ approaches the value given by the Dyakonov--Perel expression for homogenous polarization. Also shown in Fig.~\ref{fig:fig6}(c) are fits to numerical simulations of $S_z(t)$ at $(x,y)=(0,0)$ that exhibit the same features as the measured data.

\section*{Methods}

\subsection*{Experimental methods}\label{supp:expMethods}
The 2DEG investigated is confined to a 12-nm-wide (001)-oriented GaAs/AlGaAs QW placed 95$\,$nm below the surface. Asymmetric Si modulation doping provides a sufficiently strong Rashba SOI to nearly balance the Dresselhaus SOI contribution. We employ a pump-probe approach in which a circularly polarized pulse of a mode-locked laser at $\lambda=785$\,nm (full-width at half-maximum of 10\,nm) excites spin-polarized electrons into the conduction band of the QW. The spin polarization component $S_z$ along the $z\parallel[001]$ direction is measured by a probe pulse at $\lambda=799-801\,$nm using the polar magneto-optical Kerr effect~\cite{Stephens2004,Crooker2005,Meier2007}. Both pulses are directed through a single high-numeric-aperture lens placed inside the cryostat. The spatial position of the pump beam on the sample surface is scanned by controlling the angle of the incident pump beam with a Galvo mirror. In spatial maps of the spin distribution $S_z(x,y)$, the coordinates $x$ and $y$ indicate the position of the probe beam relative to the scanning pump beam. A non-linearity in the angle control was corrected to obtain calibrated values for $x$ and $y$. The pump pulse is spectrally removed with a low-pass filter prior to detection.
The focal distance is optimized to achieve a maximal spatial resolution of 2$\,\mu$m in diameter. The probe pulse is synchronized to the pump pulse and delayed with a mechanical stage by a time $t$ to monitor the time evolution of the spin polarization $S_z(t)$. The pulse lengths of the pump and the probe beam are $\approx 60\,$ps and 3$\,$ps, respectively.
Typical power intensities of pump and probe were $250$ and $50\,\mu$W, respectively, with pulses arriving at a repetition rate of 80\,MHz.
Fig.~\ref{fig:fig3}(d) includes data with the same probe power, but 100 and 30$\,\mu$W pump power. The data of Fig.~\ref{fig:fig6} have been taken at a pump and probe power of $150$ and $50\,\mu$W, respectively. All measurements were carried out at a sample temperature of 40\,K. Photo-excited electrons equilibrate rapidly ($<1$\,ps) to a Fermi distribution, however the initial electron temperature may well exceed the lattice temperature. On a time-scale smaller than 100\,ps, it is expected that these electrons are cooled by phonon emission~\cite{Ryan1984} and by scattering with the cold background electron gas. Transport measurements on etched mesa structures with Ohmic contacts yielded $n_s=5\times 10^{15}\,$m$^{-2}$  and $\mu=22\,$m$^2$/Vs. From $n_s$ we estimate a Fermi energy of 18$\,$meV above the QW ground state. The QW absorbtion edge lies at 1.54$\,$eV at 40K. In the experiment, we have to account for a finite laser spot size, leading to a finite width of the spin distribution at $t=0$. Therefore we introduced a time offset $t_0$ in the fit function $(t-t_0)^{-1}\exp(-t/\tau_s)$. This adds an uncertainty in the $\tau_s$ extracted [error bars in Fig.~\ref{fig:fig6}(c)].

\subsection*{Numerical methods}\label{supp:numMethods}

A numerical simulation of spin dynamics is performed using a two-dimensional Monte Carlo approach in which at time $t=0$, 20000 spins are distributed at coordinates $(x,y)$ with a Gaussian probability distribution of width 2$\,\mu$m. Wave numbers $(k_x, k_y)$ are uniformly distributed on the Fermi disc, and the spins $\mathbf S$ are oriented along the $z$-direction. In discrete time steps, updates of $(x,y)$ and $\mathbf S$ are calculated, treating the spin precession about the sum of $\mathbf B_\textrm{SO}+\mathbf B$ semiclassically. Scattering is accounted for by isotropically redistributing the charge carriers on the Fermi disc with scattering probability $\tau = 2 D_s/v_\textrm F^2$, where $v_\textrm F$ is the Fermi velocity. For the simulation, we used a single set of SO coefficients, $\alpha=1.7, \beta_1=2.7, \beta_3=0.7\times10^{-13}\,$eVm, and $D_s=385\,$cm$^2$s$^{-1}$ to reproduce the measured $S_z$ as shown in Figs.~\ref{fig:fig5},~\ref{fig:fig6} and~\ref{fig:fig8}. The variation of the SO coefficients with $t$, as observed in the measurement, has not been included in the simulations.


\section*{Supplementary information}

\subsection*{The effective spin-orbit magnetic field and dephasing of the persistent spin helix.}

The spin-orbit interaction for a two-dimensional electron gas confined in a (001)-grown GaAs QW is described by the three spin-orbit parameters $\alpha$, $\beta_1$, and $\beta_3$. It can be expressed as an effective spin-orbit magnetic field $\mathbf B_\textrm{SO}$, which in the coordinate system $x \parallel[110]$ and $y\parallel[1\overline10]$  is given by~\cite{Kainz2004}

\begin{equation}
\mathbf B_\textrm{SO} = \frac{2}{g \mu_B} \left(\begin{array}{c}
(\alpha+\beta_1+2\beta_3\frac{k_x^2-k_y^2}{k^2})k_y\\
(-\alpha+\beta_1-2\beta_3\frac{k_x^2-k_y^2}{k^2})k_x\end{array}\right),
\label{eq:Hamiltonian}
\end{equation}
with $\beta_1=-\gamma \langle k_z^2 \rangle$ and $\beta_3=-\gamma k^2/4$. $\langle k_z^2 \rangle$ is the expectation value of $k_z^2$ with respect to the QW ground-state envelope wave-function, $k=\sqrt{2\pi n_s}$ is the Fermi wave number of the 2DEG with sheet density $n_s$, and $\gamma$ is the Dresselhaus coupling parameter. It is convenient to write $\mathbf B_\textrm{SO}=\mathbf B^{(1)}_\textrm{SO}+\mathbf B^{(3)}_\textrm{SO}$~\cite{Lueffe2011}, with

\begin{equation}
\label{B1}
\mathbf B^{(1)}_\textrm{SO} = \frac{2 k}{g \mu_B} \left(\begin{array}{c}
(\alpha+\beta_1-\beta_3)\sin \theta\\
-(\alpha-\beta_1+\beta_3)\cos \theta\end{array}\right),
\end{equation}

and

\begin{equation}
\mathbf B^{(3)}_\textrm{SO} = \frac{2 k}{g \mu_B} \left(\begin{array}{c}
\beta_3\sin 3\theta\\
-\beta_3\cos 3\theta\end{array}\right).
\end{equation}
Here $\theta$ is the angle between $\mathbf k$ and the $x$-axis. Cross-correlations $\langle B_{\textrm{SO},x} \: k_y\rangle$ and $\langle B_{\textrm{SO},y} \: k_x\rangle$ between components of $\mathbf B_\textrm{SO}$ and components of $\mathbf k$ are responsible for the formation of helical spin modes. As $\mathbf B^{(3)}_\textrm{SO}$ disappears in these terms, the spatial pattern of the spin modes can be described by $\mathbf B^{(1)}_\textrm{SO}$ alone. In general, a PSH forms if  $\alpha=\pm (\beta_1-\beta_3)$. In our sample, $\mathbf B^{(1)}_\textrm{SO}$ mainly points along the $x$-direction, which according to Eq.~(\ref{B1}) is the case if $\alpha$ and $\beta_1-\beta_3$ have the same sign. In this situation, the helicity of the mode is determined by the sign of the correlation $\langle B_{\textrm{SO},x} \: k_y\rangle=k^2/(g\mu_B)\cdot(\alpha+\beta_1-\beta_3)$. From the shift of the PSH in an external magnetic field [Fig.~2(d)], we determine a positive helicity, i.e., a $\omega^+$ mode, and therefore $\alpha+\beta_1-\beta_3>0$.

The relaxation time $\tau_s$ of the helical mode can be expressed~\cite{Yang2010} in terms of squared components $\langle B_{\textrm{SO},x}^2\rangle$, $\langle B_{\textrm{SO},y}^2\rangle$, where $\langle...\rangle$ denotes averaging across the Fermi disc, as well as $\langle B_{\textrm{SO},x} \: k_y\rangle$. Close to the SU(2) symmetry point, i.e., for $\alpha \approx \beta_1-\beta_3$, the contribution from $\mathbf B^{(1)}_\textrm{SO}$ to $\tau_s^{-1}$ is proportional to $(\alpha-\beta_1+\beta_3)^2$ and the one from $\mathbf B^{(3)}_\textrm{SO}$ to $3\beta_3^2$ [see Eq.~(1)].


\begin{figure}[ht]
\includegraphics[width=70mm]{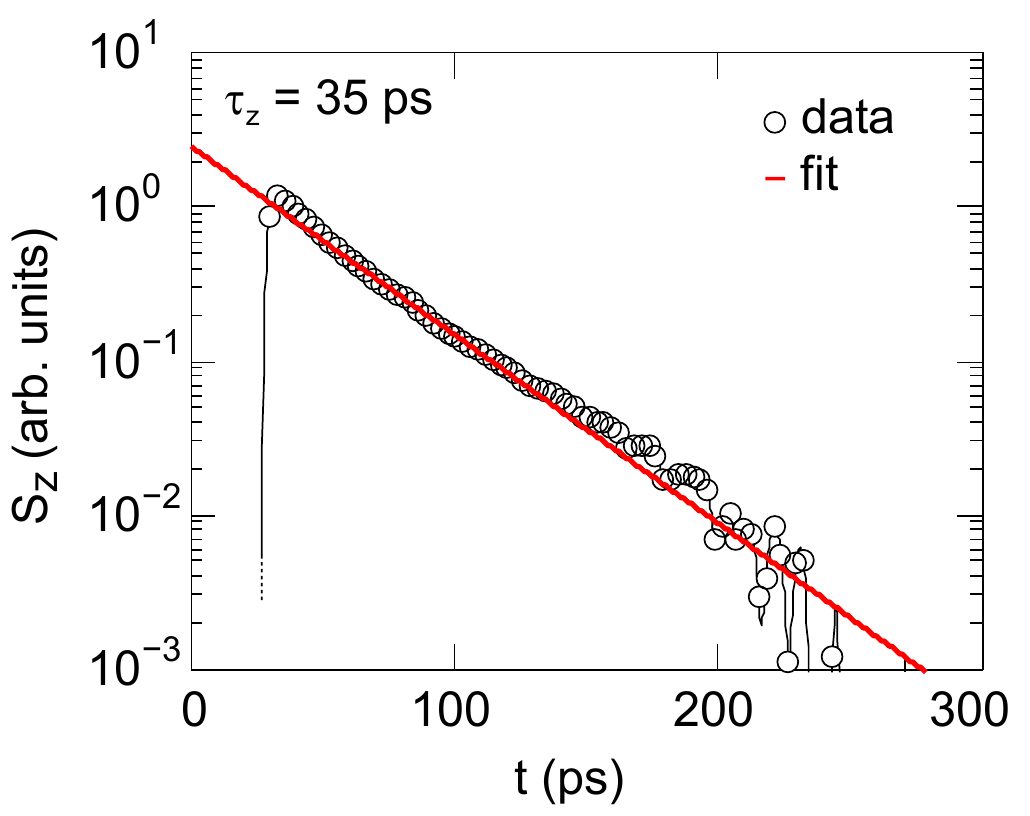}
\caption{\label{fig:fig7} \textbf{Dyakonov-Perel dephasing rate} Time trace $S_z(t)$ of spin polarization for a measurement with $20$-$\mu$m-wide laser spots. In this case, spatial spin-orbit correlations are averaged out and $S_z(t)$ decays with the Dyakonov-Perel dephasing rate. }
\end{figure}

\subsection*{Dyakonov-Perel dephasing rate}\label{supp:dephasingrates}
In the diffusive limit, where the scattering length is much smaller than the spin-orbit length and if spins perform a random walk on the Bloch sphere (Dyakonov Perel regime), the expected decay rate for a spin polarization along the $z$-direction is~\cite{Kainz2004}

\begin{equation}\label{eq:DyakonovPerel}
\tau_{z}^{-1} = 8 D_s m^2 \hbar^{-4} \big[\alpha^2 + (\beta_1 - \beta_3)^2 + \beta_3^2 \big].
\end{equation}

If the spatial correlations of the PSH are averaged out, $\tau_s$ is given by Eq.~(\ref{eq:DyakonovPerel}) instead of Eq.~(1). Using 20-$\mu$m-wide laser spots, we measure $\tau_{z}\approx 35\,$ps at 30~K (Fig.~\ref{fig:fig7}). This is in perfect agreement with the value obtained from Eq.~(\ref{eq:DyakonovPerel}) using $\alpha=2.3\times10^{-13}\,$eVm, $\beta_1-\beta_3=2.6\times10^{-13}\,$eVm, $\beta_3=0.7\times10^{-13}\,$eVm and $D_s=385\,$cm$^2$s$^{-1}$.

\begin{figure}[h!t]
\includegraphics[width=75mm]{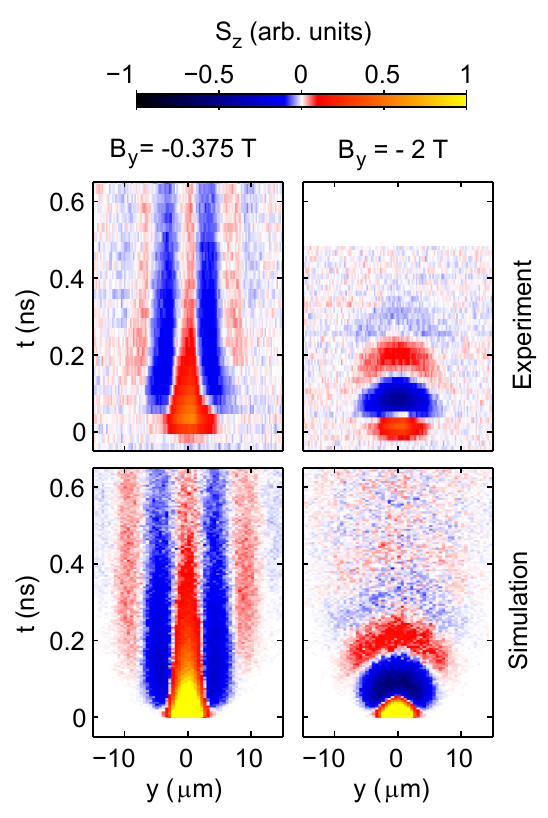}
\caption{\label{fig:fig8} \textbf{Detuning from the PSH regime.} Map of $S_z(y,t)$ for $B_y=-0.375$ and $-2$\,T. Top panels: Experimental data (same data as shown in Fig.~6). Bottom panels: Numerical simulation.}
\end{figure}

\subsection*{Numerical simulation data}

To compare experiment and simulation, we present the numerically simulated $S_z(y,t)$-maps for $\mathbf{B}\parallel y$ that correspond to the experimental data shown in Fig.~\ref{fig:fig6}. Simulation and experiment agree remarkably well in both the $\mathbf{B}_{\textrm{SO}}$- and the $\mathbf{B}$-dominated regime (Fig.~\ref{fig:fig8}).


\section*{Acknowledgments}
We would like to acknowledge financial support from the Swiss National Science Foundation through NCCR Nano and NCCR QSIT, as well as valuable discussions with Rolf Allenspach, Klaus Ensslin and Yuansen Chen.


\end{document}